\documentclass[pra,aps,twocolumn]{revtex4-1}
\usepackage{graphicx}
\usepackage{float}
\usepackage{amssymb}

\usepackage{hyperref}
\usepackage{amsmath}
\usepackage{graphicx}
\usepackage{braket}
\usepackage{mathtools}

\begin{document}

\pagestyle{headings}
\title{Intrinsic anharmonic effects on the phonon frequencies and effective spin-spin interactions in a quantum simulator made from trapped ions in a linear Paul trap}
\author{M. McAneny}
\affiliation{Department of Physics, Georgetown University, 37th and O Sts. NW,
Washington, DC 20057, USA}

\author{J. K. Freericks}
\affiliation{Department of Physics, Georgetown University, 37th and O Sts. NW,
Washington, DC 20057, USA}


\date{\today}

\begin{abstract}
The Coulomb repulsion between ions in a linear Paul trap give rise to anharmonic terms in the potential energy when expanded about the equilibrium positions. We examine the effect of these anharmonic terms on the accuracy of a quantum simulator made from trapped ions.  To be concrete, we consider a linear chain of $\text{Yb}^{171+}$ ions stabilized close to the zigzag transition.  We find that for typical experimental temperatures, frequencies change by no more than a factor of $0.01\%$ due to the anharmonic couplings.  Furthermore, shifts in the effective spin-spin interactions (driven by a spin-dependent optical dipole force) also tend to be small for detunings to the blue of the transverse center-of-mass frequency.  However, detuning the spin interactions near other frequencies can lead to nonnegligible anharmonic contributions to the effective spin-spin interactions.  We also examine an odd behavior exhibited by the harmonic spin-spin interactions for a range of intermediate detunings, where nearest neighbor spins with a larger spatial separation on the ion chain interact more strongly than nearest neighbors with a smaller spatial separation.
\end{abstract}

\maketitle

\section{Introduction}
In 1982, Richard Feynman opened the field of quantum simulation when he proposed that
quantum simulators can  be employed in order to study the evolution and interactions of complex quantum mechanical systems~\cite{feynman}.  It is only recently that ion trap quantum simulators have demonstrated success in engineering model spin-systems in both one dimensional and two dimensional lattices of trapped
 ions~\cite{two-ion,Kim1,Kim2,Edwards,Islam,John,John2,Blatt,Islam2,Richerme,Blatt2,Senko}.  Starting with the demonstration of the effective spin interaction between two ions~\cite{two-ion}, it was shown that larger numbers of ions interact with well-defined Ising spin exchange~\cite{Kim1}, which can show frustration~\cite{Kim2,Edwards}, and can be scaled to approximate the thermodynamic phase transition~\cite{Islam}. Penning trap experiments~\cite{John,John2}, showed that the same concepts can be exptended to hundreds of ions trapped in a rotating triangular lattice. The idea of stroboscopic quantum simulation has also been shown~\cite{Blatt}.  Recently, systems have been scaled up to 18 ions~\cite{Islam2} and properties of dynamics and excited states have been examined via Lieb-Robinson-like studies of correlation growth~\cite{Richerme,Blatt2} and spectroscopy of excited states~\cite{Senko}. These ion-trap systems work well due to their long decoherence times, scalability, and ability to be precisely controlled experimentally.

As the precision of these experiments grows, one needs to examine perturbations of these systems that
carry them away from the simplest ideal.  In addition, as the system sizes grow, it becomes increasingly 
difficult to fully cool the systems down to low temperature as Raman sideband cooling becomes
more complex and difficult to carry out.  In addition, it is often only the phonon modes that are to be driven that are cooled below the Doppler limit, the phonons in other spatial directions are often left at the Doppler limit, which can have them with tens to hundreds of quanta excited. 

Anharmonic effects enter into an oscillator when the period of the oscillation depends on its amplitude. In solid state physics, anharmonic effects are well known in causing lattices to (typically) expand as they are heated. Another way of describing this behavior is that as anharmonic terms are considered, they break
the simple picture of free normal modes that oscillate at their own independent frequencies into a coupled oscillator system that can have its periods change, that can have resonantly enhanced dissipations, and that can excite quanta in the coupled modes. It is impossible to completely remove anharmonic effects from an ion trap, even if the trapping potentials can be made purely harmonic, because there is an intrinsic anharmonicity that arises due to the Coulomb interaction between the ions. In this work, we investigate whether such anharmonic effects are likely to cause inaccuracies in a quantum simulation.

Anharmonic effects have been considered previously for linear Paul traps. James showed how one can determine the coupling tensors that arise due to the anharmonic nature of the Coulomb interaction and how one can use those couplings to resonantly dissipate energy from one mode to the other modes via optical-mixing-like effects~\cite{James}. This transfer of energy from one mode to another was investigated experimentally in a two-site chain~\cite{Roos}.  The effect of anharmonicities in either the potentials or the Coulomb interaction were investigated to see how phonon frequencies shift due to the occupancy of other phonon modes~\cite{Wineland}. In this work, we focus on how the intrinsic anharmonicity affects the phonon frequencies and how those, in turn, affect the Ising spin exchange couplings.

The remainder of this paper is organized as follows:
In Sec.tion II, we discuss the formalism for determining anharmonic effects, numerical results follow in Section III,
and we conclude in Section IV. Details of the anharmonic coupling tensors appear in the appendices.

\section{Theoretical Formulation}
We consider a chain of ions in a linear Paul trap, which uses a combination of static and frequency-dependent fields in order to trap ions.  The precise behavior of this system often includes micromotion due to the time-dependent fields, but it is well described by a static pseudopotential, when the ion equilibrium positions lie at the nulls of the potential energy surface. We provide our analysis under the assumption that the static pseudopotential approach is accurate for describing the motion of the ions in the trap.

The potential energy describing such a system of $N$ ions includes both a term describing the Coulomb interaction between each pair of ions, and a term related to the spring energy of each ion along the $z$-axis (which will be the axis of longitudinal alignment for the ions). The dimensionless potential is then given by:
\begin{equation}\label{pot}
V=\frac{1}{2}\sum_{\alpha=x,y,z}\sum_{i=1}^N \beta_\alpha^2 x_{ i\alpha}^2+\sum_{\substack{i,j=1\\j\not=i}}^N \frac{1}{|{\bf r}_{i}-{\bf r}_{j}|}
\end{equation}
where the potential has been scaled by $m\omega_z^2l_0^2$, and the dimensionless ion coordinates ${\bf r}_i=(x_{ix},x_{iy},x_{iz})$ have been renormalized by a characteristic length $l_0=[k Z^2e^2/(m\omega_z^2)]^{1/3}$ with $k$ the Coulomb coupling constant, $Z$ the charge on the ion, $e$ the charge of an electron, and $m$ the mass of the ion. In addition, we have $\beta_\alpha=\omega_\alpha/\omega_z$ where $\omega_\alpha$ is the trapping frequency in the $\alpha$ direction.  We consider the case with $\beta_{x}=\beta_{y}\gg\beta_z=1$, which gives rise to a one-dimensional chain for the ions, if the number of ions $N$ lies below a critical value.

From Eq.~(\ref{pot}), the equilibrium positions are readily found numerically by using nonlinear optimization routines to find where the potential is a minimum and the force vanishes~\cite{James}.  Then, by expanding the potential to fourth order in the coordinates of the ions about their equilibrium positions, one obtains the Hamiltonian written in the phonon creation/annihilation operator basis as follows:
\begin{eqnarray}\label{Hamilt}
\mathcal{H}&=&\sum_{a=1}^{3N}\varepsilon_a \bigg(\hat{a}_a^\dagger \hat{a}_a+\frac{1}{2}\bigg)\\
&+&\sum_{\substack{a,b,c=1}}^{3N}B_{a,b,c}(\hat{a}_a+\hat{a}_a^\dagger)(\hat{a}_b+\hat{a}_b^\dagger)(\hat{a}_c+\hat{a}_c^\dagger)\nonumber\\
&+&\sum_{\substack{a,b,c,d=1}}^{3N}C_{a,b,c,d}(\hat{a}_a+\hat{a}_a^\dagger)(\hat{a}_b+\hat{a}_b^\dagger)(\hat{a}_c+\hat{a}_c^\dagger)(\hat{a}_d+\hat{a}_d^\dagger)\nonumber
\end{eqnarray}
where the scaled normal-mode (phonon) energies satisfy $\varepsilon_a=\hbar \omega_\alpha/(m\omega_z^2 l_0^2)$, and the explicit values for the cubic and quartic coupling tensors $B$ and $C$ are given in the appendices. The roman subscript denotes the specific normal mode, which is indexed from 1 to $3N$.

At this stage, it is appropriate to treat these higher order terms as a perturbation to the harmonic Hamiltonian because they should correspond to small corrections to the potential when the ion remains close to its equilibrium position. The third-order term creates no first-order shift to the energy spectrum, as it contains an odd number of creation/annihilation operators.  Therefore, a second-order perturbation expansion is required for that term.  On the other hand, the second-order correction due to the quartic term is considered negligible as it contains a $C^2$ term, which is much less than the $C$ or $B^2$ terms.  Then, using time-independent Rayleigh-Schr\"odinger perturbation theory, the anharmonic shifts can be calculated to first order in $C$ and second order in $B$~\cite{Wineland}.

Before proceeding, it is relevant to note that the anharmonic frequency shifts of the center-of-mass modes (both transverse and longitudinal) can both be found to identically vanish through fourth order.  This was shown explicitly to third order~\cite{James}, and can be immediately generalized to all orders, because the center-of-mass mode decouples from all other motion when the trap potential is purely harmonic and the inter-ion forces satisfy Newton's third law~\cite{dehmelt,dubin}. Hence the center-of-mass frequency it fixed at the respective trap frequency.

In general, the shifts in frequency (scaled by the transverse center-of-mass frequency, $\omega_{CM}$) can then be written as a function of the occupation number of each mode as:
\begin{equation}
\frac{\Delta \omega_a(n_a,\{n_b\})}{\omega_{CM}}=\frac{\Delta E(n_a+1,\{n_b\})-\Delta E(n_a,\{n_b\})}{ \varepsilon_{CM}}
\label{eq: freq}
\end{equation}
where CM denotes the center-of-mass.
This is an intuitive definition for the anharmonic frequency shifts, since it is the energy shift caused by adding one more phonon to the system with unperturbed frequency $\omega_a$ (when there are already $n_a$ phonons in that mode)~\cite{Wineland}.

The ions in the trap often have an internal hyperfine structure which can be mapped onto Ising spin variables. For the Yb$^{171+}$ ion, one usually takes the clock states as the ``spin up'' and ``spin down'' states. This internal (spin) degree of freedom can be coupled to the motional degrees of freedom by applying a spin-dependent optical dipole force.  This is usually done by applying red and blue detuned laser beams ($\omega\pm\mu$) on top of a carrier beam at $\omega$. By considering the AC Stark effect caused by these beams, and factorizing the resulting evolution operator, one can realize a spin-dependent optical dipole force on the incident ion~\cite{Porras,Monroe}.  In this way, the motional degrees of freedom of the system are coupled to the spin degrees of freedom, generating the following Ising Hamiltonian:
\begin{equation}
\mathcal{H}_{Ising}=\sum_{i,j}J_{i,j}(t)\sigma_i^x\sigma_j^x,
\end{equation}
with exchange coefficients that depend on time.
The time-independent piece of the spin-spin couplings is~\cite{Monroe}
\begin{equation}
J_{i,j}=\frac{F_O^2}{4m}\sum_{a=1}^N\frac{b_i^{x a}b_j^{x a}}{\mu^2-\omega_a^2}
\end{equation}
where $F_O$ is the magnitude of the spin-dependent optical dipole force and $b_i^{x a}$ is the $i^{\text{th}}$ ion's component of the transverse phonon eigenvector corresponding to the $\omega_a$ mode. In this summation, we only take the modes that lie in the direction of the driving force, which is typically the $x$ direction.

It is difficult to extend this derivation to include the cubic and quartic phonon mode coupling terms in the phonon Hamiltonian, because the operator factorization of the evolution operator becomes much more complicated (see, for example, Ref.~\onlinecite{Wang}, which shows how to factorize the evolution operator and describes the problems that arise from noncommuting operators).  Therefore, rather than calculating these complicated terms, it is assumed that these terms are small because the ions do not deviate far from their equilibrium positions.  Accordingly, we work in the quasi-harmonic approximation, for which the only change in the formalism for the $J_{i,j}$'s is that we replace the harmonic frequency with the shifted anharmonic frequency found in Eq. (\ref{eq: freq}).

Note that the average occupation number of each phonon mode can be shown to be $ n_a=[\text{exp}(\hbar \omega_a/k_B T)-1]^{-1}$ when it is in thermal equilibrium at a temperature $T$.  Then, instead of calculating the anharmonic frequency shifts (and thus the anharmonic spin-spin interactions) in terms of the occupation numbers, one can write the average occupation numbers as a function of temperature, and thus $\Delta \omega_a = \Delta \omega_a(T)$. Specifically, there are two temperature limits relevant to current experimental efforts.  The first regime is the Doppler cooling limit for all modes, where the temperature reached is on the order of a few hundred microKelvin.  The second temperature regime is Doppler cooling plus sideband cooling on the transverse modes.  For our purposes, the sideband cooling will lead to effectively zero occupation of the transverse modes, but the longitudinal modes remain at the Doppler limit temperature.

\section{Numerical Results}
In this section, we present numerical examples to illustrate how anharmonic couplings affect the frequencies and spin-spin interactions between ions in the linear Paul trap.  The parameters we use reflect typical parameters used in current experimental efforts.  The longitudinal trapping frequency is $\omega_z=500\text{ kHz}$, and the transverse trapping frequencies are given by $\beta_x=\beta_y=10$. Furthermore, we consider a trap with 24 ions arranged in it, which, for the above parameters, is the maximum number of ions without a zigzag transition (unstable modes).  The trapped ions are $\text{Yb}^{171+}$.  Note that for these parameters, the Doppler cooling limit is on the order of $k_B T\approx\hbar \omega_{CM}$ for the transverse center of mass frequency. This implies the longitudinal modes should have on the order of one to ten quanta excited at the Doppler cooling limit.

\subsection{Frequency Shifts}
\begin{figure}
\hspace*{-.1 in}
\includegraphics[width=3.5in]{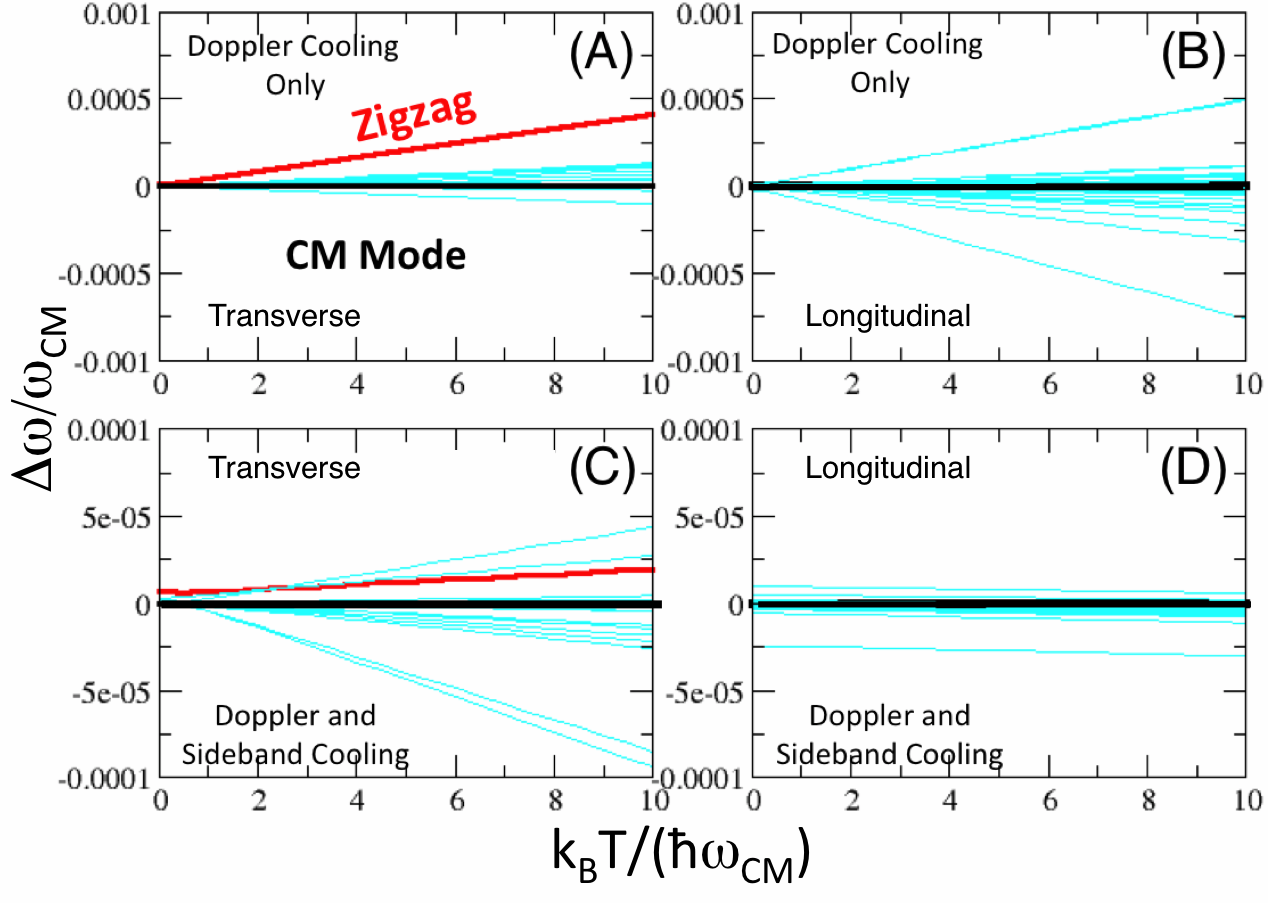}
\caption{\label{fig: fshifts} (Color online). Anharmonic frequency shifts as a function of the temperature of the phonon modes measured relative to the transverse center-of-mass phonon frequency $\omega_{\rm CM}$. A plots the frequency shifts of the transverse modes in the Doppler cooling limit, where both the transverse and longitudinal phonons are at temperature $T$.  B plots the shifts of the longitudinal modes in the Doppler cooling limit.  C shows the shifts of transverse  modes with Doppler and sideband cooling (zero transverse phonon occupation, so the transverse phonons are in the ground state, while the longitudinal ones are at temperature $T$).  D shows the shifts of longitudinal modes with Doppler and sideband cooling (zero transverse phonon occupation). }
\end{figure}

First, we examine the effects of the anharmonicities on the frequencies of the modes. Figure~\ref{fig: fshifts} shows both the anharmonic shifts of the longitudinal and transverse frequencies for Doppler cooling only, and for Doppler cooling with sideband cooling of the transverse modes.  

A few trends are immediately noticeable.  First of all, the shifts remain smaller than the order of $10^{-4}\omega_{CM}$ for the case of Doppler cooling only (top panels), and smaller than the order of $10^{-5}\omega_{CM}$ for the case with sideband cooling also (bottom panels).  This suggests not only that the anharmonic frequency shifts are relatively small, but also that sideband cooling can suppress these shifts another order of magnitude.  Furthermore, it is worth noting that, as anticipated analytically, the shifts for both the transverse and longitudinal center-of-mass modes are zero (black lines in Figure~\ref{fig: fshifts}).  In experiment, with a phonon frequency on the order of a MHz, the anharmonic shift would be below the order of 100~Hz for Doppler cooling and below 10~Hz for Doppler plus side band cooling. We expect effects on the order of 100~Hz to be experimentally observable, but smaller shifts will be difficult to see, and are unlikely to affect other aspects of the experiments.

The frequency shift curves are most nearly linear in temperature.  This may be surprising considering that the expression for $\Delta E$ has a quadratic dependency on the occupation numbers of different modes (cf. Appendix A).  However, when we take $\Delta E(n_a+1,\{n_b\})-\Delta E(n_a,\{n_b\})$, the quadratic dependencies cancel out, and since $n_a$ is roughly linear with temperature except at extremely small temperatures, we find that the frequency shifts are linear with temperature.

\begin{figure}[htb!]
\hspace*{-.1 in}
\includegraphics[width=3.4in]{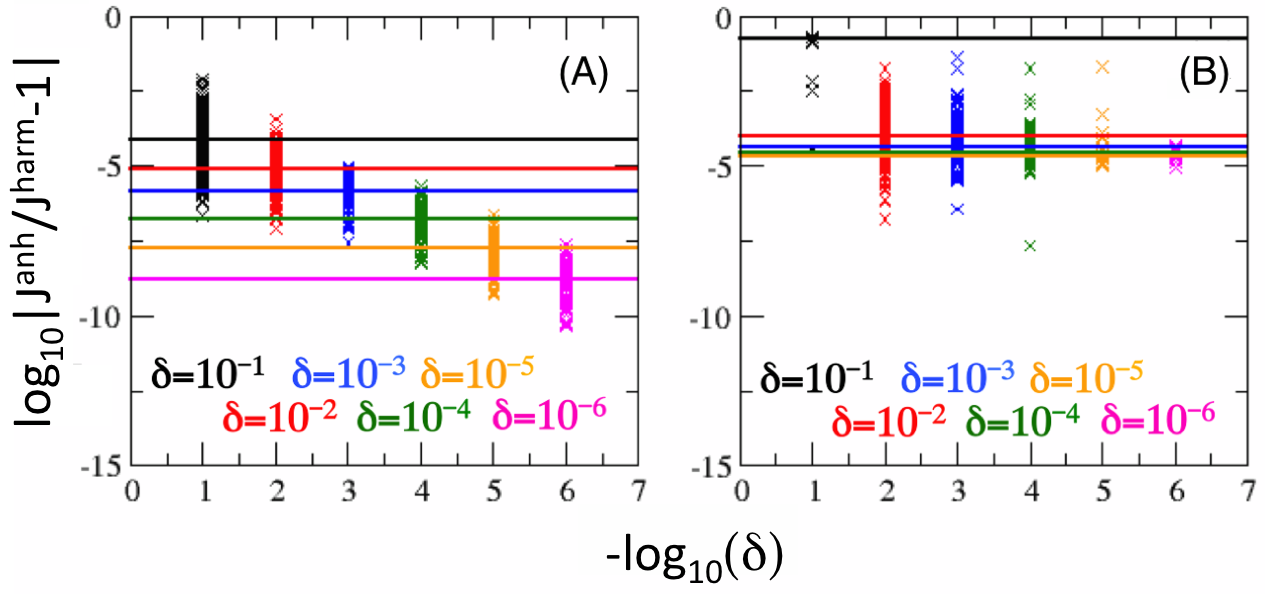}
\caption{\label{fig: JvsDet} (Color online.) Panel A shows the proportional shifts in the spin-spin interactions for detunings above the center-of-mass mode.  Panel B shows the same for detunings above the fifth smallest transverse frequency mode. Both plots are shown for $\delta=10^{-1}$, $\delta=10^{-2}$, $\delta=10^{-3}$, $\delta=10^{-4}$, $\delta=10^{-5}$, and $\delta=10^{-6}$. 
The symbols show the shifts for all ($24\times 23/2=276$) spin-spin interactions $J_{ij}$. The solid horizontal lines show the average shift. }
\end{figure}

Next, we consider the shifts in the effective static spin-spin interactions.  Particularly, we discuss how the spin-spin interactions are affected when $\mu=\omega_a(1+\delta)$, for which we say that the spin-spin interactions are detuned by $\delta$ above mode $a$. Figure~\ref{fig: JvsDet} plots the proportional change in the spin-spin interactions between the harmonic and anharmonic Hamiltonians for Doppler cooling only.  Panel A shows the shifts in the spin-spin interactions for detunings above the transverse  center-of-mass mode, while panel B has detunings above the fifth smallest transverse frequency.

Clearly, panel A exhibits negligible shifts, especially for detunings smaller than $\delta=10^{-2}$.  For the $\delta=10^{-1}$ detuning (black curve), the spin interactions are shifted by as much as 1\%, although the average is more toward the order of 0.01\%.  Furthermore, as the detuning decreases by an order of magnitude, so do the shifts.  In this vein, the smallest detuning of $\delta=10^{-6}$ (magenta curve) shifts by a factor of approximately $10^{-9}$.  These shifts are obviously negligible.

On the other hand, panel B, which shows detuning above the fifth smallest frequency mode, exhibits quite different behavior.  First of all, the shifts are on average about 10\% for the largest detuning (black curve).  Furthermore, for detunings as small as $10^{-5}$, there are still shifts by a factor of 1\%, although the average shift is more toward 0.01\% for these smaller detunings.  Then in general the anharmonic effects on the spin-spin interactions are not negligible, even for the small detunings.

\begin{figure}[thb!]
\hspace*{-.1 in}
\includegraphics[width=3.65in]{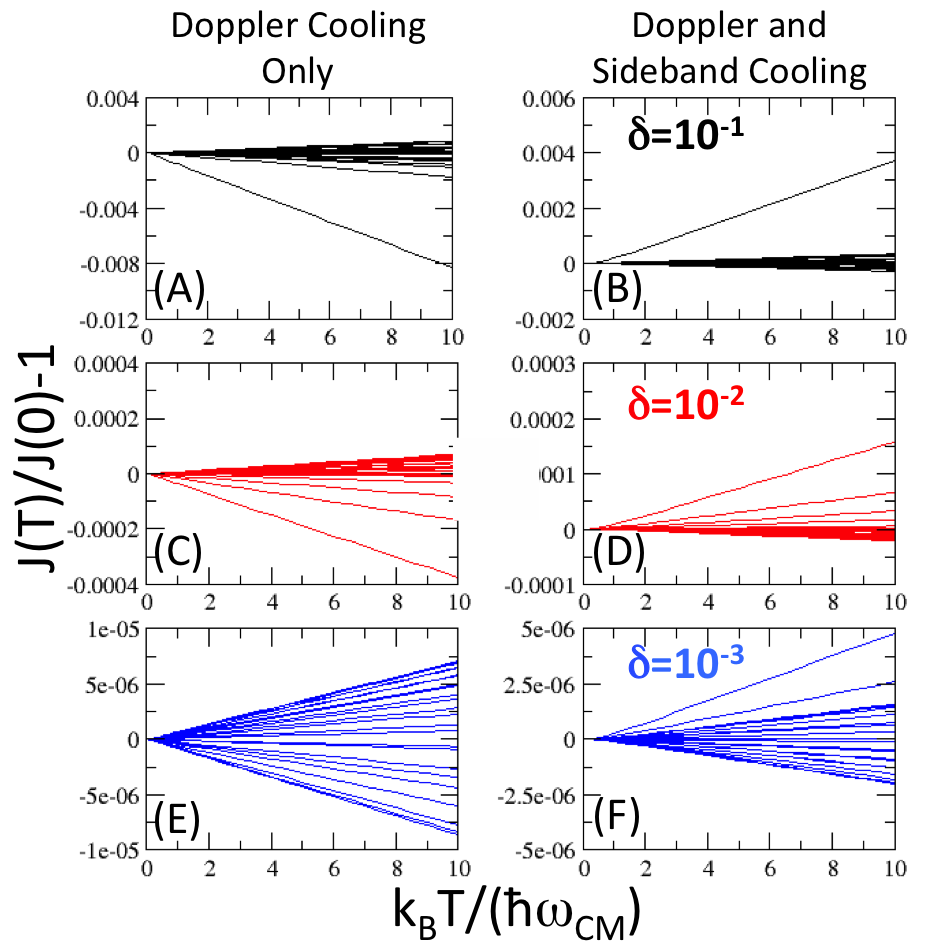}
\caption{\label{fig: JT}(Color online.)  Dependence of the spin-spin interactions detuned above the center of mass mode as a function of the system's temperature (Doppler cooling only on the left, Doppler plus sideband cooling on the right).  The spin-spin interactions plotted are $J_{1j}$, that is the spin interactions between the left-most ion in the chain with every other ion and we use three different detunings corresponding to the different colors.  The most significant trend in this plot is that the shifts are nearly linear as a function of temperature. Furthermore, in panels A and B, and even in panels C and D, there are clear outliers that shift much more than the other lines.  As expected, these lines plot the spin interactions between the ions on either side of the chain; the shifts are proportionally large because the spin-spin interactions between distant sites are quite small to begin with, hence the effect of these shifts, even though they appear to be large, are likely to be rather small on the dynamics of the system. }
\end{figure}

This phenomenon can be attributed to the fact that $\Delta\omega_{CM}=0$.  Since the shift in the center-of-mass frequency is zero, the changes in $J_{i,j}$ that one might expect to see due to the $(\mu^2-\omega_{CM}^2)^{-1}$ term are largely suppressed.  However, due to nonzero shifts in frequency for other modes, detuning above other modes makes the spin-spin interactions more sensitive to anharmonic effects. 

Figure~\ref{fig: JT} shows how the spin-spin interactions between the first ion and every other ion shift as a function of temperature for detunings above the center of mass.  The spin-spin interactions appear roughly linear as a function of temperature.  Furthermore, particularly for the larger detunings, there are some outliers that shift significantly more than the others.  These shifts are for the interactions between the first ion and the farthest away ion.  Since those interactions themselves are small for large detunings and far distances, even small changes in the spin interactions will change the interaction between these sites by a significant amount.  In fact, even for the $10^{-1}$ detuning, many of the interactions change by no more than a factor of $10^{-7}$.

\begin{figure}[h!]
\hspace*{-.2 in}
\includegraphics[width=3.5in]{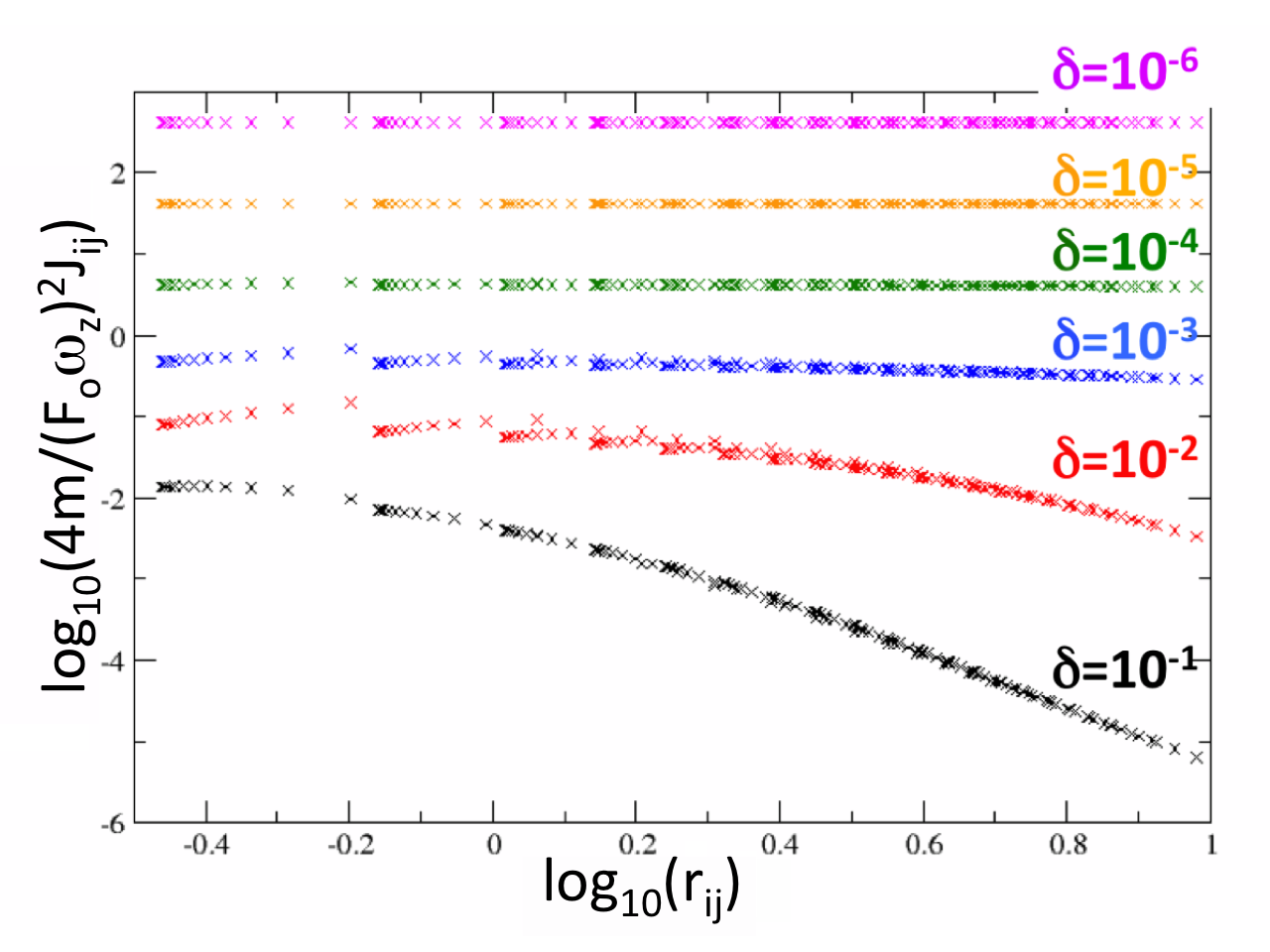}
\caption{\label{fig: harmJ} (Color online.) Harmonic spin-spin interactions as a function of the distance between the interacting spins.  Note that for intermediate detuning $\delta = 10^{-2}$ and $\delta = 10^{-3}$ curves (red and blue) and for small separations between spins, there can be increasing spin interactions for increasing separation between the ions.  This suggests that ions towards either end of the chain interact more strongly with their neighbors than those in the middle of the chain do, a surprising result since ions in the middle of the chain are closer together. For larger distances, the spin-spin couplings become approximate power laws, as expected.  }
\end{figure}

Finally, we found an interesting trend in the harmonic spin-spin interactions that warrants further discussion.  Figure~\ref{fig: harmJ} shows the harmonic spin-spin couplings $J_{i,j}$'s as a function of the distance between the interacting spins.  As a whole, the graph is in fact fairly typical: the smallest detunings produce the largest spin interactions (that are approximately constant), while the largest detunings cause smaller spin-spin interactions that fall off like $J=r^{-3}$.  However, if one looks closely at the curves corresponding to intermediate detunings $\delta=10^{-3}$ and $\delta=10^{-2}$ (the blue and red curves respectively), then one notices local regions at small distances, for which the spin interactions {\it increase} as the distance between the interacting ions increases.  Essentially, since nearest neighbor ions on the inside of the chain are compressed closer together than neighbors on the outer edges of the chain, this suggests that ions toward the outside of the chain interact more strongly with their neighbors than the inner ions do with their neighbors.  This effect is also readily seen for smaller detunings, although the scale of Figure~\ref{fig: harmJ} does not easily show this.
Such spin-spin couplings could potentially allow for interesting types of spin models to be examined, since the couplings change character---initially growing with distance and then decaying, and they also show that it is not always true that the spin-spin couplings can be described by a simple power law behavior, as is often assumed. It is likely that this behavior is due to the fact that the phonon modes with frequencies close to the center-of-mass mode (such as the tilt mode) have larger phonon displacements for the ions furthest from the center, than phonon modes farther from the center-of-mass phonon frequency (such as the zig-zag mode).

\section{Conclusion and Discussion}
In this work, we have treated the potential of the linear Paul trap to fourth order in order to consider the effects of anharmonic couplings on the normal mode frequencies and spin-spin interactions between trapped ions. We find that the frequency shifts are small (on the order of $10^{-4}\omega_{CM}$) when only Doppler cooling is utilized, and another order of magnitude smaller when sideband cooling is also implemented.  Furthermore, we find that spin-spin interactions that are detuned above the center of mass mode change proportionally very little, which is a consequence of the fact that $\Delta\omega_{CM}=0$.  However, for spin interactions detuned near other modes, the anharmonic couplings can have an appreciable effect.  Finally, we find that the spins toward the ends of the ion chain counterintuitively interact more strongly with their neighbors than do the spins on the inside of the chain for a wide range of detunings to the blue of the transverse center-of-mass mode. The mechanism behind this phenomenon is currently unclear, however, and will require further analysis for a fuller understanding. But it is likely related to the fact that the modes closest to the center-of-mass mode have the largest relative phonon displacements for ions furthest from the center of the trap, as compared to those that have phonon frequencies further from the center-of-mass frequency, where the ion motion of the normal modes are dominated by ions closer to the center of the trap.

\acknowledgments
We acknowledge John Bollinger for suggesting this problem to us and for valuable discussions.
This work was supported by the National Science Foundation under grant number PHY-1314295.
 J. K. F. also acknowledges support from the McDevitt bequest at Georgetown University.

\appendix
\section{Derivation of the anharmonic coupling Hamiltonian}\label{hamiltonianDerivation}
In order to generate the Hamiltonian given by Eq.~\ref{Hamilt}, we first expand the potential to fourth order about the equilibrium positions of the ions, and then we transform this result from the ion position basis to the phonon mode basis.  Let $\mathcal{H}_0$ denote the harmonic Hamiltonian for the system.  Then,
\begin{align}
\mathcal{H}&=\mathcal{H}_0+\frac{1}{6}\sum_{\substack{\alpha,\beta,\gamma\\=x,y,z}}\sum_{\substack{i,j,k=1}}^N \frac{\partial^3 V}{\partial x_{\alpha i} \partial x_{\beta j}\partial x_{\gamma k}}\Bigg|_0 \epsilon_{\alpha i}\epsilon_{\beta j} \epsilon_{\gamma k}\nonumber\\
&+\frac{1}{24}\sum_{\substack{\alpha,\beta,\gamma,\delta\\=x,y,z}}\sum_{\substack{i,j,k,l=1}}^N \frac{\partial^4 V}{\partial x_{\alpha i} \partial x_{\beta j}\partial x_{\gamma k}\partial x_{\delta l}}\Bigg|_0 \epsilon_{\alpha i}\epsilon_{\beta j} \epsilon_{\gamma k}\epsilon_{\delta l}
\end{align}
where $\epsilon_{\alpha i}$ is given by $\epsilon_{\alpha i}=x_{\alpha i}-x_{\alpha i}^0$ for equilibrium positions $x_{\alpha i}^0$, and where the partial derivatives are evaluated at the equilibrium positions. 

Now define:
\begin{equation}
\tilde B_{\alpha i,\beta j,\gamma k}=\frac{\partial^3 V}{\partial x_{\alpha i} \partial x_{\beta j}\partial x_{\gamma k}}\Bigg|_0
\end{equation}
and
\begin{equation}
\tilde C_{\alpha i, \beta j, \gamma k, \delta l}= \frac{\partial^4 V}{\partial x_{\alpha i} \partial x_{\beta j}\partial x_{\gamma k}\partial x_{\delta l}}\Bigg|_0.
\end{equation}
These tensors can of course be solved for by taking the third and fourth derivatives of the potential (given by equation~\ref{pot}), and then evaluating them at equilibrium.  However, the calculations themselves are quite tedious, so the lengthy algebra is omitted.  The final results for $\tilde B$ and $\tilde C$ are shown in appendix~\ref{fullExpressions}.  

Now, we must change to the phonon-mode basis.  In the phonon-mode basis, the harmonic Hamiltonian is
given in terms of the creation/annihilation operators, and the displacement from equilibrium $\epsilon_{\alpha i}$ is replaced by the phonon displacement operator $X_a^{\alpha}$ which must be summed over all phonon modes and weighted by the normal mode eigenvectors to yield the total displacement. The Hamiltonian becomes
\begin{widetext}
\begin{equation*}
\mathcal{H}=\mathcal{H}_0^\text{phon}+\frac{1}{6}\sum_{\substack{\alpha,\beta,\gamma\\=x,y,z}}\sum_{\substack{i,j,k=1}}^N\sum_{\substack{a,b,c=1}}^{3N}b_a^{\alpha  i}b_b^{\beta j}b_c^{\gamma k}\tilde B_{\alpha i,\beta j, \gamma k}X_aX_bX_c
+\frac{1}{24}\sum_{\substack{\alpha,\beta,\gamma,\delta\\=x,y,z}}\sum_{\substack{i,j,k,l=1}}^N \sum_{\substack{a,b,c,d=1}}^{3N}b_a^{\alpha i} b_b^{\beta j} b_c^{\gamma k} b_d^{\delta l}\tilde C_{\alpha i, \beta j, \gamma k, \delta l}X_aX_bX_cX_d.
\end{equation*}
\end{widetext}
where the symbol $b_a^{\alpha i}$ denotes the eigenvector of the harmonic phonon Hamiltonian for the $a$th mode, showing the displacement of the $i$th ion in the $\alpha$th direction~\cite{James}.
Define the following ladder operators:
\begin{equation}
\hat a_a=\sqrt{\frac{1}{2\varepsilon_a}}\bigg(\frac{\omega_a}{\omega_z}X_a+iP_a\bigg)\nonumber
\end{equation}
\begin{equation}
\hat a_a^\dagger=\sqrt{\frac{1}{2\varepsilon_a}}\bigg(\frac{\omega_a}{\omega_z}X_a-iP_a\bigg),
\end{equation}
which satisfy the canonical commutation relations.
Then, by expressing the Hamiltonian in terms of the ladder operators, we obtain~\cite{James}:
\begin{align}\label{eq: Hamilt}
\mathcal{H}&=\sum_{a=1}^{3N}\varepsilon_a \bigg(\hat{a}_a^\dagger \hat{a}_a+\frac{1}{2}\bigg)\nonumber\\
&+\sum_{\substack{a,b,c=1}}^{3N}B_{a,b,c}(\hat{a}_a+\hat{a}_a^\dagger)(\hat{a}_b+\hat{a}_b^\dagger)(\hat{a}_c+\hat{a}_c^\dagger)\nonumber\\&+\sum_{\substack{a,b,c,d=1}}^{3N}C_{a,b,c,d}(\hat{a}_a+\hat{a}_a^\dagger)(\hat{a}_b+\hat{a}_b^\dagger)(\hat{a}_c+\hat{a}_c^\dagger)(\hat{a}_d+\hat{a}_d^\dagger)
\end{align}
where
\begin{align}
B_{a,b,c}=\frac{1}{6}\bigg(\frac{\hbar}{2 m l_0^2}\bigg)^{3/2}&(\omega_a \omega_b \omega_c )^{-1/2}\nonumber\\&
\sum_{\substack{i,j,\\k=1}}^{3N}\sum_{\alpha,\beta,\gamma}\tilde{B}_{\alpha i,\beta j, \gamma k}b_a^{\alpha i}b_b^{\beta j}b_c^{\gamma k}
\end{align}
and
\begin{align}
C_{a,b,c,d}=\frac{1}{24}\bigg(\frac{\hbar}{2 m l_0^2}\bigg)^2&(\omega_a \omega_b \omega_c \omega_d)^{-1/2}\nonumber\\&\sum_{\substack{i,j,\\k,l=1}}^{3N}\sum_{\substack{\alpha,\beta,\\\gamma, \delta}}\tilde{C}_{\alpha i,\beta j, \gamma k, \delta l}b_a^{\alpha i}b_b^{\beta j}b_c^{\gamma k}b_d^{\delta l}.
\end{align}
Note that $B$ and $C$ are simply the coefficients of the third and fourth order potential terms ($\tilde{B}$ and $\tilde{C}$ are in the ion position basis) written in the phonon basis with some constants absorbed.

\section{Expressions for $\tilde B$ and $\tilde C$}\label{fullExpressions}

\begin{widetext}
We now evaluate the exact expressions for the cubic and quartic phonon mode coupling tensors in terms of the equilibrium positions, as follows:
\begin{align*}
\alpha=x,y: \tilde B_{\alpha i,\alpha j,zk}&=
\begin{cases}
-3 \displaystyle\sum_{\substack{\nu=1\\\nu\not=i}}^N\frac{\text{sgn}(\nu-i)}{|x^0_{z\nu}-x^0_{zi}|^{4}}, &i=j=k.\\
3\displaystyle\frac{\text{sgn}(j-i)}{|x^0_{zj}-x^0_{zi}|^{4}},&\text{two indices are $i$, other is $j$},\\
0,&i\not=j,j\not=k,k\not=i.
\end{cases}\nonumber\\
\tilde B_{zi,zj,zk}&=
\begin{cases}
6 \displaystyle\sum_{\substack{\nu=1\\\nu\not=i}}^N\frac{\text{sgn}(\nu-i)}{|x^0_{z\nu}-x^0_{zi}|^{4}}, &i=j=k.\\
-6\displaystyle\frac{\text{sgn}(j-i)}{|x^0_{zj}-x^0_{zi}|^{4}},&\text{two indices are $i$, other is $j$},\\
0,&i\not=j,j\not=k,k\not=i.
\end{cases}\nonumber\\
\alpha=x,y: \tilde C_{\alpha i,\alpha j,\alpha k,\alpha l}&=
\begin{cases}
 \displaystyle\sum_{\substack{\nu=1\\\nu\not=i}}^N\frac{9}{|x^0_{z\nu}-x^0_{zi}|^5}, &i=j=k=l,\\
\displaystyle\frac{-9}{|x^0_{zi}-x^0_{zj}|^5},&\text{three indices $i$, other is $j$},\\
\displaystyle\frac{9}{|x^0_{zi}-x^0_{zj}|^5},& \text{two indices $i$, other two are $j$},\\
0,& \text{otherwise}.
\end{cases}\nonumber\\
\alpha,\beta=x,y; \alpha\not=\beta: \tilde C_{\alpha i,\alpha j,\beta k,\beta l}&=
\begin{cases}
\displaystyle\sum_{\substack{\nu=1\\\nu\not=c}}^N\frac{3}{|x^0_{z\nu}-x^0_{zi}|^5}, &i=j=k=l,\\
\displaystyle\frac{-3}{|x^0_{zi}-x^0_{zj}|^5},&\text{three indices $i$, other is $j$},\\
\displaystyle\frac{3}{|x^0_{zi}-x^0_{zj}|^5},&\text{two indices $i$, other two are $j$},\\
0,& \text{otherwise}.
\end{cases}\nonumber\\
\alpha=x,y: \tilde C_{\alpha i,\alpha j,z k,z l}&=
\begin{cases}
\displaystyle\sum_{\substack{\nu=1\\\nu\not=i}}^N\frac{-12}{|x^0_{z\nu}-x^0_{zi}|^5}, &i=j=k=l,\\
\displaystyle\frac{12}{|x^0_{zi}-x^0_{zj}|^5},&\text{three indices $i$, other is $j$},\\
\displaystyle\frac{-12}{|x^0_{zi}-x^0_{zj}|^5},& \text{two indices $i$, other two are $j$},\\
0,& \text{otherwise}.
\end{cases}\nonumber\\
\tilde C_{zi,zj,zk,zl}&=
\begin{cases}
\displaystyle\sum_{\substack{\nu=1\\\nu\not=i}}^N\frac{24}{|x^0_{z\nu}-x^0_{zi}|^5}, &i=j=k=l,\\
\displaystyle\frac{-24}{|x^0_{zi}-x^0_{zj}|^5},& \text{three indices $i$, other is $j$},\\
\displaystyle\frac{24}{|x^0_{zi}-x^0_{zj}|^5},&\text{two indices $i$, other two are $j$},\\
0,& \text{otherwise}.
\end{cases}
\end{align*}
\end{widetext}

\section{Derivation of anharmonic energy shifts via nondegenerate perturbation theory}
To solve for the energy shifts, we consider the first- and second-order Rayleigh-Schr\"odinger corrections for the quartic and cubic perturbations, respectively.  The anharmonic energy shifts are then be given by the following:
\begin{align}
\Delta E(\{n_a\})&=\prescript{}{0}{\bra {\{n_a\}}} V^{(4)} \ket{\{n_a\}}_0\nonumber\\&+\sum_{\{m_a\}\not=\{n_a\}}\frac{|\prescript{}{0}{\bra {\{m_a\}}} V^{(3)} \ket{\{n_a\}}_0|^2}{E_n^0-E_m^0}
\end{align}
Recall that the first-order correction for the third-order potential term is 0 since there are an odd number of creation/annihilation operators.  Furthermore, recall that we ignore the second-order correction for the fourth-order potential term as it is negligible.

Keeping in mind that $E_n^0=\sum_{a=1}^N\varepsilon_a\bigg(n_a+\frac{1}{2}\bigg)$, and that:
\begin{equation}
V^{(3)}=\sum_{\substack{a,b,\\c=1}}^{3N}B_{a,b,c}(\hat a_a^\dagger+\hat a_a)(\hat a_b^\dagger+\hat a_b)(\hat a_c^\dagger+\hat a_c)
\end{equation}
\begin{equation}
V^{(4)}=\sum_{\substack{a,b,\\c,d=1}}^{3N}C_{a,b,c,d}(\hat a_a^\dagger+\hat a_a)(\hat a_b^\dagger+\hat a_b)(\hat a_c^\dagger+\hat a_c)(\hat a_d^\dagger+\hat a_d)
\end{equation}
one can directly solve for the anharmonic energy shifts. After some tedious algebra, the final expression for the anharmonic energy shifts becomes
\begin{widetext}
\begin{align*}
\Delta E(\{n_\alpha\})&=3\sum_{\alpha=1}^{3N}\bigg[(2n_\alpha^2+2n_\alpha+1)C_{\alpha\alpha\alpha\alpha}+2(2n_\alpha+1)\sum_{\beta\not=\alpha}^{3N}(2n_\beta+1)C_{\alpha\alpha\beta\beta}\bigg]\nonumber\\
&+\bigg(\frac{\omega_z^2 l_0^2 m}{2\hbar}\bigg)\bigg[-\sum_{\alpha=1}^{3N}B_{\alpha\alpha\alpha}^2\frac{30n_\alpha^2+30n_\alpha+11}{\omega_\alpha}-18\sum_{\alpha=1}^{3N}\sum_{\beta\not=\alpha}^{3N}B_{\beta\beta\alpha}B_{\alpha\alpha\alpha}\frac{(2n_\beta+1)(2n_\alpha+1)}{\omega_\alpha}\nonumber\\
&+9\sum_{\alpha=1}^{3N}\sum_{\beta\not=\alpha}^{3N}B_{\beta\beta\alpha}^2\bigg(\frac{-4\omega_\beta(2n_\beta+1)(2n_\alpha+1)}{4\omega_\beta^2-\omega_\alpha^2}+\frac{2(n_\beta^2+n_\beta+1)}{4\omega_\beta^2-\omega_\alpha^2}-\frac{(2n_\beta+1)^2}{\omega_\alpha}\bigg)\nonumber\\
&-18\sum_{\alpha=1}^{3N}\sum_{\beta\not=\alpha}^{3N}\sum_{\substack{\gamma\not=\beta\\\gamma\not=\alpha}}^{3N}B_{\beta\beta\alpha}B_{\gamma\gamma\alpha}\frac{(2n_\beta+1)(2n_\gamma+1)}{\omega_\alpha}
+36\sum_{\alpha=1}^{3N}\sum_{\beta\not=\alpha}^{3N}\sum_{\substack{\gamma\not=\beta\\\gamma\not=\alpha}}^{3N}B_{\alpha\beta\gamma}^2\nonumber\\
&\times\bigg(-\frac{(\omega_\alpha+\omega_\beta)(1+n_\alpha+n_\beta)(2n_\gamma+1)}{(\omega_\alpha+\omega_\beta)^2-\omega_\gamma^2}+\frac{\omega_\gamma(1+n_\alpha+n_\beta+2n_\alpha n_\beta)}{(\omega_\alpha+\omega_\beta)^2-\omega_\gamma^2}
+\frac{(\omega_\alpha-\omega_\beta)(n_\alpha-n_\beta)(2n_\gamma+1)}{(\omega_\alpha-\omega_\beta)^2-\omega_\gamma^2}\nonumber\\
&+\frac{\omega_\gamma(n_\alpha+n_\beta+2n_\alpha n_\beta)}{(\omega_\alpha-\omega_\beta)^2-\omega_\gamma^2}\bigg)\bigg].
\end{align*}
This formula is used to evaluate the anharmonic shifts in the phonon frequencies in the main text. Note that these formulas are at most quadratic in the phonon occupation numbers, and become linear when differences are taken with neighboring occupancies.
\end{widetext}

\end{document}